\documentclass[article,reprint,amsmath,amssymb,longbibliography,superscriptaddress]{revtex4-1}
\usepackage{graphicx}
\usepackage{dcolumn}
\usepackage{bm}
\usepackage{hyperref}
\usepackage{breakurl}
\usepackage{multirow}
\usepackage{enumitem}
\usepackage[pagewise]{lineno}
\setcitestyle{super}

\begin{document}

\title{Pressure-induced superconductivity in topological semimetal NbAs$_{2}$}

\author{Yupeng Li}
\thanks{Equal contributions}
      \affiliation{Department of Physics, Zhejiang University, Hangzhou 310027, China}

\author{Chao An}
\thanks{Equal contributions}
      \affiliation{Anhui Province Key Laboratory of Condensed Matter Physics at Extreme Conditions, High Magnetic Field Laboratory, Chinese Academy of Sciences, Hefei 230031, China}

\author{Chenqiang Hua}
\thanks{Equal contributions}
     \affiliation{Department of Physics, Zhejiang University, Hangzhou 310027, China}

\author{Xuliang Chen}
     \affiliation{Anhui Province Key Laboratory of Condensed Matter Physics at Extreme Conditions, High Magnetic Field Laboratory, Chinese Academy of Sciences, Hefei 230031, China}

\author{Yonghui Zhou}
     \affiliation{Anhui Province Key Laboratory of Condensed Matter Physics at Extreme Conditions, High Magnetic Field Laboratory, Chinese Academy of Sciences, Hefei 230031, China}

\author{Ying Zhou}
     \affiliation{Anhui Province Key Laboratory of Condensed Matter Physics at Extreme Conditions, High Magnetic Field Laboratory, Chinese Academy of Sciences, Hefei 230031, China}

\author{Ranran Zhang}
     \affiliation{Anhui Province Key Laboratory of Condensed Matter Physics at Extreme Conditions, High Magnetic Field Laboratory, Chinese Academy of Sciences, Hefei 230031, China}

\author{Changyong Park}
      \affiliation{HPCAT, Geophysical Laboratory, Carnegie Institution of Washington, Argonne, Illinois 60439, USA}

\author{Zhen Wang}
      \affiliation{Department of Physics, Zhejiang University, Hangzhou 310027, China}
       \affiliation{State Key Laboratory of Silicon Materials, Zhejiang University, Hangzhou 310027, China}

\author{Yunhao Lu}
      \affiliation{State Key Laboratory of Silicon Materials, Zhejiang University, Hangzhou 310027, China}

\author{Yi Zheng}
      \affiliation{Department of Physics, Zhejiang University, Hangzhou 310027, China}
      \affiliation{Collaborative Innovation Centre of Advanced Microstructures, Nanjing University, Nanjing 210093, China}

\author{Zhaorong Yang}
      \email{zryang@issp.ac.cn}
      \affiliation{Anhui Province Key Laboratory of Condensed Matter Physics at Extreme Conditions, High Magnetic Field Laboratory, Chinese Academy of Sciences, Hefei 230031, China}
      \affiliation{Collaborative Innovation Centre of Advanced Microstructures, Nanjing University, Nanjing 210093, China}

\author{Zhu-An Xu}
      \email{zhuan@zju.edu.cn}
      \affiliation{Department of Physics, Zhejiang University, Hangzhou 310027, China}
      \affiliation{State Key Laboratory of Silicon Materials, Zhejiang University, Hangzhou 310027, China}
      \affiliation{Collaborative Innovation Centre of Advanced Microstructures, Nanjing University, Nanjing 210093, China}

\date{\today}

\begin{abstract}

Topological superconductivity with Majorana bound states, which are critical to implement non-abelian quantum computation, may be realized in three-dimensional semimetals with nontrivial topological feature, when superconducting transition occurs in the bulk. Here, we report pressure-induced superconductivity in a transition-metal dipnictide NbAs$_{2}$. The emergence of superconductivity is not accompanied by any structural phase transition up to the maximum experimental pressure of 29.8 GPa, as supported by pressure-dependent synchrotron x-ray diffraction and Raman spectroscopy. Intriguingly, the Raman study reveals rapid phonon mode hardening and broadening above 10 GPa, in coincident with the superconducting transition. Using first-principle calculations, we determine Fermi surface change induced by pressure, which steadily increases the density of states without breaking the electron-hole compensation. Noticeably, the main hole pocket of NbAs$_{2}$ encloses one time-reversal-invariant momenta of the monoclinic lattice, suggesting NbAs$_{2}$ as a candidate of topological superconductors.

\end{abstract}

\maketitle

\noindent\textbf{INTRODUCTION}

\noindent There have been various proposals in search of quasiparticle
excitations of Majorana fermions (MFs) in solids, which are the
subject of both fundamental research and error-tolerant
topological quantum computing
\cite{Quantumcomputation_Nayak_RMP,Majorana_Wilczek_NatPhys}.
Superconductor-topological insulator (TI) heterostructures turn
the surface Dirac fermions of topological insulators (TIs) into
$p$-wave-like Cooper pairs \cite{Proximity_FuLiangPRL08}. Majorana
zero modes by this superconducting proximity approach have been
observed in vortices by scanning tunnelling microscopy in
Bi$_{2}$Te$_{3}$/NbSe$_{2}$ heterostructures
\cite{Majorana_JiaJF_PRL15,Majorana_SunHH_PRL}.  Superconducting
proximity effects may also create MFs in semiconducting nanowires
with strong spin-orbital coupling
\cite{Majorana_Sarma_PRL10A,Majorana_Sarma_PRL10B}, or in
ferromagnetic atomic chains \cite{Majorana_Yazdani_Science14}.
Alternatively, chiral Majorana edges states are expected in
two-dimensional (2D) chiral topological superconductors,
consisting of a topological insulator in proximity to an s-wave
superconductor \cite{ChiralSC_PRB_QiXL10,
ChiralSCMeasurement_PRB_ZhangSC11, ChiralSCQHE_PRB_ZhangSC15,
ChiralSC_Science_WangKL17}. It is intriguing that superconducting
phase transitions are widely observed in many topological
materials when high pressure is applied, such as type II Weyl
semimetals of MoTe$_{2}$\cite{MoTe2_YanpQi_NatC} and
WTe$_{2}$\cite{WTe2_DFKang_NatC,WTe2_XCPan_NatC}, Dirac semimetals
of Cd$_{3}$As$_{2}$\cite{Cd3As2_HeLP2016QuantMater} and
ZrTe$_{5}$\cite{ZrTe5HP_zhouYH_PANS}, and topological insulators
of Bi$_{2}$Se$_{3}$\cite{Bi2Se3_Kevin_PRL},
Bi$_{2}$Te$_{3}$\cite{Bi2Te3_zhangJL_PNAS}, and
Sb$_{2}$Te$_{3}$\cite{Sb2Te3_ZhuJ_ScientificR}. Recently
tip-induced superconductivity has attracted much attention because
it could offer a new platform to study topological
superconductivity (TSC) in Dirac \cite{Cd3As2PointT_WangHe_NatM}
and Weyl semimetals \cite{tipWeylSC_WangH_SB17}. Charge carrier
doping is another effective method to induce TSC in topological
insulators such as in Cu$_{x}$Bi$_{2}$Se$_{3}$
\cite{CuxBi2Se3_HorYS_PRL} and Sr$_{x}$Bi$_{2}$Se$_{3}$
\cite{SrxBi2Se3_Shruti_PRB15}. The experimental observations
strongly suggest the feasibility to realize TSC
\cite{TopologicalSC_ZSC_PRL09,CuxBi2Se3_FuL_PRL} in novel
topological materials with non-trivial topological features.
However, before the emergence of superconductivity (SC), most of
these topological compounds undergo structural phase transitions,
which usually change the topological states.

Very recently, transition-metal dipnictides of the MPn$_{2}$
family
\cite{TaSb2_YukL_PRB,NbAs2family_YupengLi_arXiv,NbAs2_Shenb_PRB,TaAs2_FangZ_arXiv,TaAs2_LuoYK_SR,TaAs2_ShuangJ_PRB,TaAs2_XiaTL_PRB},
where M represents Mo, Nb or Ta atom, and Pn is As or Sb atom,
respectively, generate a wide interest as a new prototype of
topological semimetals with extremely large magnetoresistance
(XMR). When spin orbital coupling (SOC) is included, the band
anticrossing in MPn$_{2}$ between the M-$d_{xy}$ and
M-$d_{x^{2}-y^{2}}$ orbitals along the $I-L-I'$ direction are
fully gapped \cite{TaSb2_WZ_Lifshitz}, resulting in weak
topological insulator invariants of
$\mathbb{Z}_2=[0;111]$\cite{NbAs2family_YupengLi_arXiv,TaAs2_LuoYK_SR,NbAs2familiyculc_caochao_PRB}.
Another interesting point in the MPn$_{2}$ family is that magnetic
field could induce Weyl points \cite{NbAs2Weyl_GreschD_NJP17}.
With its monoclinic lattice (Space group No. 12), which is
normally stable under very high pressure
\cite{ZrTe5HP_zhouYH_PANS,Cd3As2_HeLP2016QuantMater}, it is
tempting to study the physical properties of MPn$_{2}$ under high
pressure.

\begin{figure*}[!thb]
\begin{center}
\includegraphics[width=7in]{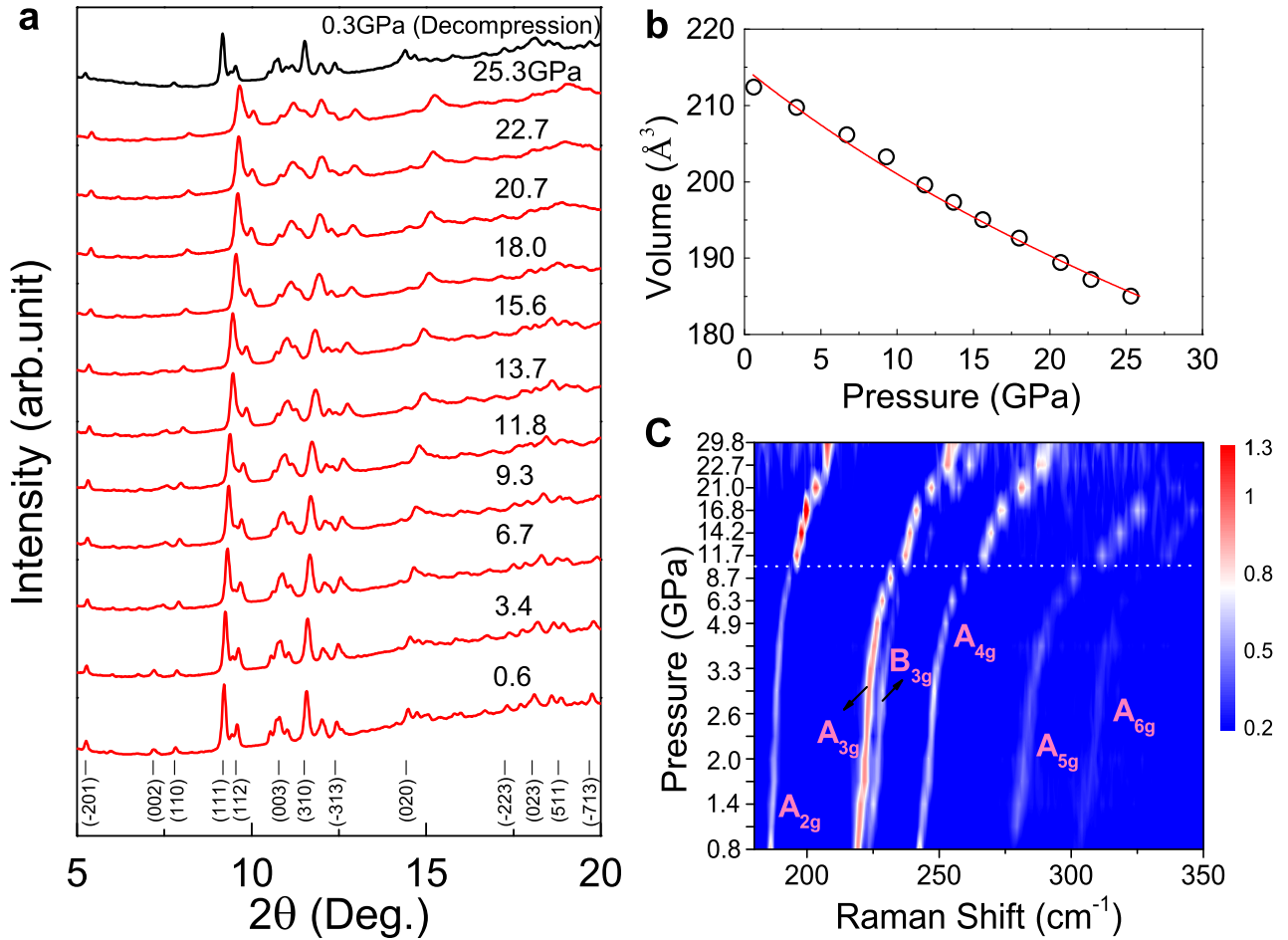}
\end{center}
\caption{\label{Fig1} High pressure structure of NbAs${_2}$. (\textbf{a}) X-ray diffraction patterns of NbAs$_{2}$ at various pressures display no structural phase transition. (\textbf{b}) Unit-cell volume decreases with increasing pressure, while the volume diminishes by 13\% at 25.3 GPa. (\textbf{c}) The Raman shifts at various pressures show the hardening and broadening of Raman scattering peaks.}
\end{figure*}

In this work, we report pressure-dependent transport measurement
and structure evolution in NbAs$_{2}$, using the
diamond-anvil-cell (DAC) technique to continuously tune the
electronic structure. Superconducting transition has been
successfully observed at 12.8 GPa with a critical temperature
($T_\text{c}$) of 2.63 K. Further increase in pressure gradually
suppresses $T_\text{c}$, which disappears completely at 27.9 GPa
when bad metal behavior dominates below the helium temperature.
Using high-pressure X-ray diffraction and Raman scattering, we
confirm that there is no structural phase transition up to the
maximum experimental pressure of 29.8 GPa, which implies that the
topological state maybe remain undisturbed under high pressure.
Intriguingly, the electron-hole compensation in NbAs$_{2}$ changes
little even in the superconducting region, as suggested by
first-principle calculations, although the electron-hole
compensation under pressure is not so perfect. Instead, the
dwindling of XMR and the emergence of superconductivity are
characterized by continuous Fermi surface change and gradual
growth in density of state. Our results thus indicate the
coexistence of XMR and SC. The growing density of states is
usually beneficial to the occurrence of SC. Moreover, the Cooper
pair formation in NbAs$_{2}$ may also be correlated to the
enhanced electron-phonon coupling under high pressure, since the
Raman studies reveal significant phonon mode hardening and
broadening above 10 GPa. It is noteworthy that, in the SC region,
the main hole pocket of NbAs$_{2}$ encloses the
time-reversal-invariant (TRI) momenta $M$. These suggest
NbAs$_{2}$ as a candidate of TRI topological superconductors.

\vspace{3ex}

\noindent\textbf{RESULTS AND DISCUSSION}

\begin{figure*}[!thb]
\begin{center}
\includegraphics[width=6.5in]{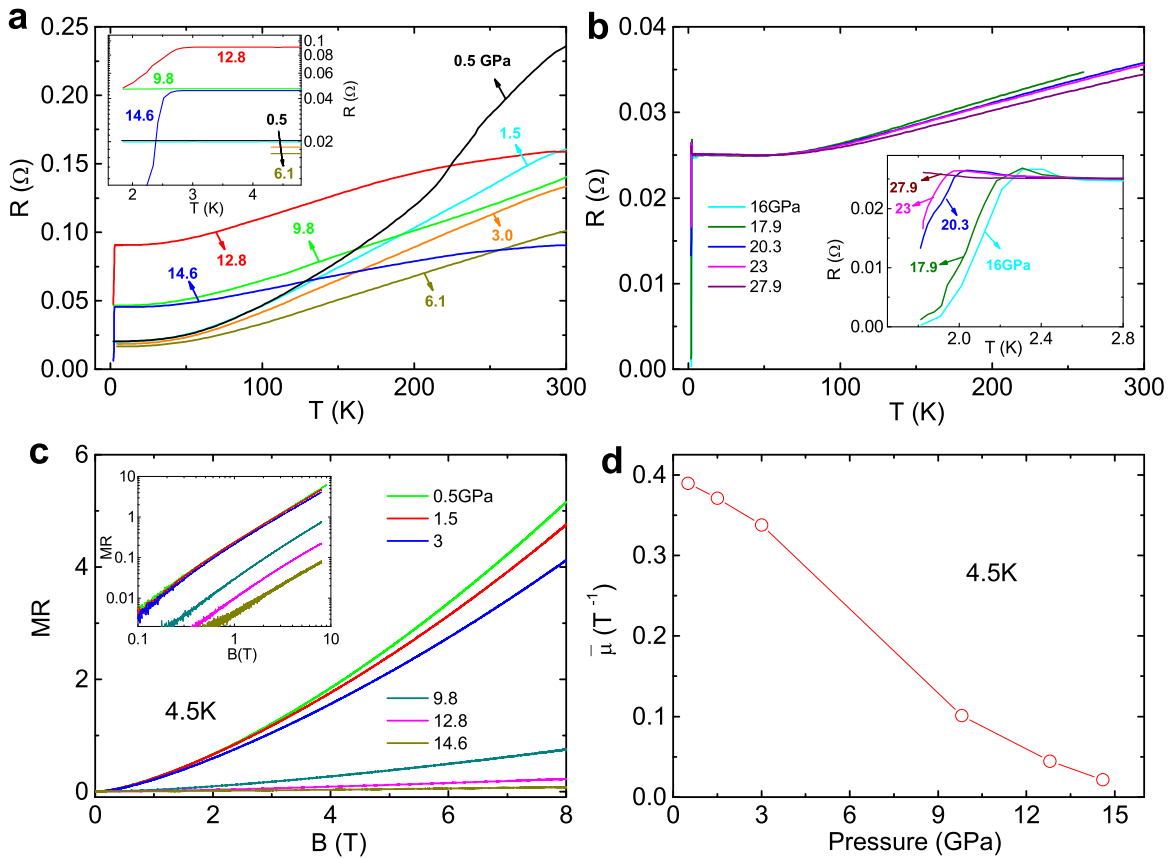}
\end{center}
\caption{\label{Figure2} Pressure-induced superconductivity in
single-crystal NbAs$_{2}$. (\textbf{a}) Temperature-dependent
resistance is measured from 0.5 GPa to 14.6 GPa and
superconducting transition appears at 12.8 GPa. The enlarged view
of low temperature resistance is shown in the inset. (\textbf{b})
The resistance as a function of temperature at different pressures
between 16 GPa and 27.9 GPa is displayed with a enlarged plot
around the superconducting transition in the inset. (\textbf{c})
Magnetoresistance (MR=(R(8T)-R(0T))/R(0T)) at 4.5 K is suppressed
slowly as pressure increases from 0.5 GPa to 14.6 GPa. The MR at
12.8 GPa and 8 T is nearly 22\%, which is still very large.
(\textbf{d}) Average mobility $\bar{\mu}$ obtained by fitting to
the formula $\text{MR}=(\bar{\mu}B)^{1.43}$ decreases monotonously
with increasing pressure.}
\end{figure*}

\noindent Pressure-dependent structure analysis

\noindent The structure evolution of NbAs$_{2}$ under pressure is determined by synchrotron radiation-based high-pressure X-Ray diffraction (XRD) and Raman spectroscopy, as shown in Fig.\ref{Fig1}. The XRD data are refined using the Rietveld method, and the Bragg peaks are well indexed by the space group $C2/m$ for both the 0.6 GPa and 25.3 GPa data (See Supplementary Fig.S1). Within 30 GPa, the pressure-induced lattice changes are reversible. In Fig.\ref{Fig1}a, we show the XRD results of NbAs$_{2}$ when the DAC is slowly decompressed from 29.8 GPa to 0.3 GPa. The data are essentially identical to the bottom curve of 0.6 GPa. In Fig.\ref{Fig1}b, the high pressure dependent unit-cell volumes can be fitted by the third$-$order Birch-Murnaghan equation of state\cite{BMfitting1947PR}, and no distinct drop of the unit-cell volume is observed with the increase of pressure, which suggests stability of this structure. The monoclinic lattice is also manifested in pressure-dependent Raman scattering measurements, showing six fingerprinting peaks between 180 and 350 cm$^{-1}$, namely $A_\text{2g}$, $A_\text{3g}$, $B_\text{3g}$, $A_\text{4g}$, $A_\text{5g}$, and $A_\text{6g}$, respectively (see Fig.\ref{Fig1}c). It is intriguing that Raman spectroscopy shows distinctive enhancement of $A_\text{2g}$, $A_\text{3g}$, $B_\text{3g}$, $A_\text{4g}$ and $A_\text{5g}$ above 10 GPa, when the data are normalized by the $A_\text{2g}$ peak intensity. The distinct broadening in $A_\text{2g}$, $A_\text{4g}$ and $A_\text{5g}$, which could be an indication of stronger electron-phonon coupling or pressure-enhanced defect scattering \cite{RamanNbAs2_ZhangQM_PRB16}, is also discernable above 10 GPa.

\vspace{3ex}

\noindent Resistivity measurements

\noindent Fig.\ref{Figure2}a and \ref{Figure2}b summarize the resistance vs temperature ($R-T$) characteristics of NbAs$_{2}$ at various pressures from 0.5 GPa to 27.9 GPa. Below 9.8 GPa, NbAs$_{2}$ shows typical metallic behaviour, which is characterized by monotonic increase in resistance with increasing temperature, and exhibits a lower residual-resistivity-ratio (RRR) with enhancement of pressure. The trend is reversed at 9.8 GPa, when the whole $R-T$ curve is upshifted by $\sim0.03\, \Omega$ compared to 6.1 GPa. Superconductivity with a transition temperature of 2.63 K emerges at 12.8 GPa, but the resistance does not drop to zero even at 1.8 K (Fig.\ref{Figure2}a). Further increase in pressure surprisingly starts to suppress $T_\text{c}$, although the SC transition becomes sharper, and zero-resistance behaviour is observed at 16 GPa. Noticeably, the zero-resistance transition is accompanied by a positive resistance kink, which is very sharp and only extends within a temperature range of 0.2 K. Subsequently, the superconducting transition is gradually suppressed when the positive resistance kink becomes broader and its onset shifts to lower $T$. Above 23 GPa, the positive resistance kink completely dominates.

\begin{figure*}[!thb]
\begin{center}
\includegraphics[width=6.5in]{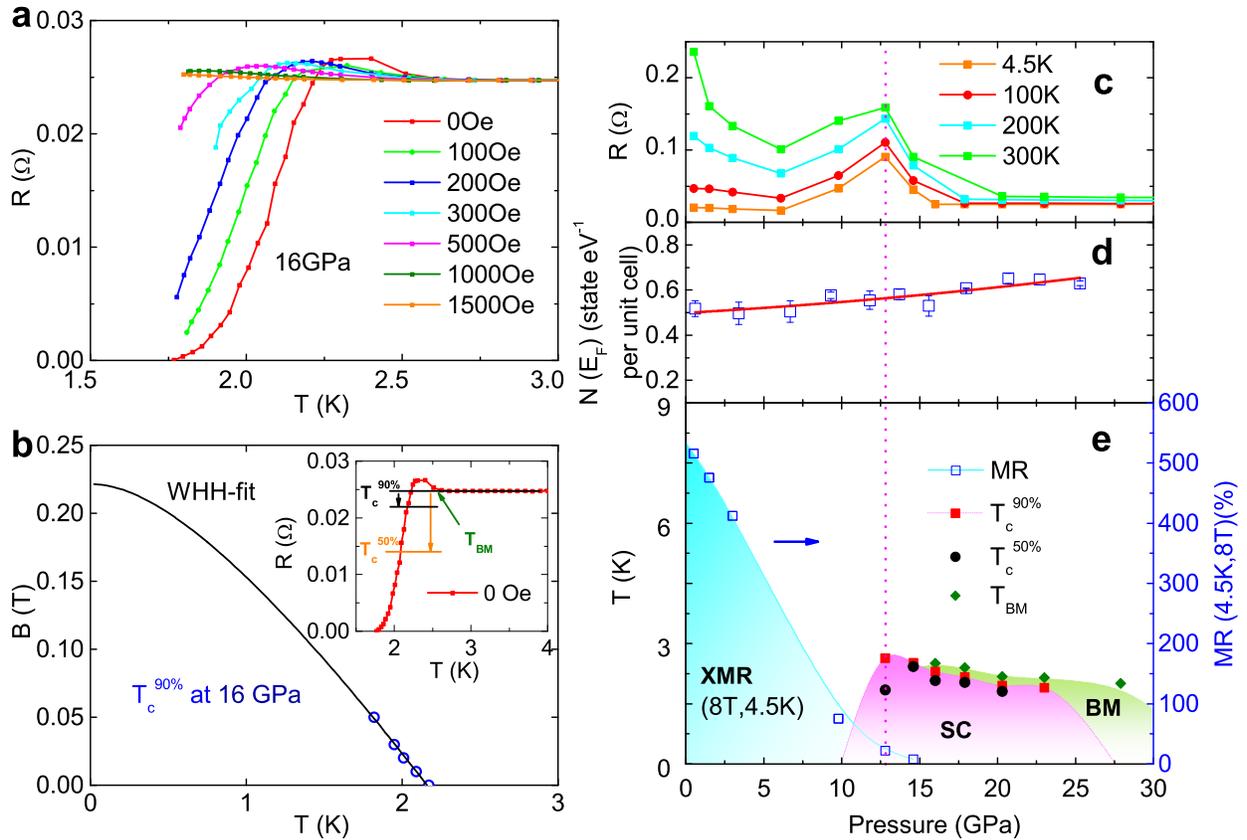}
\end{center}
\caption{\label{Figure 3} Determination of upper critical field
for superconductivity and temperature-pressure phase diagram of
NbAs$_{2}$. (\textbf{a}) Temperature-dependent resistance at
different magnetic fields is measured at 16 GPa. (\textbf{b})
H$_{c2}$ as a function of temperature and its WHH fitting are
displayed. The inset shows the definition of different
superconducting transition temperatures and transition temperature
of bad metal. (\textbf{c}) Resistance at 4.5 K, 100 K, 200 K and
300 K in the normal state at different pressures shows the peak
near 12.8 GPa where superconducting transition emerges.
(\textbf{d}) Calculated density of states of NbAs$_{2}$ at the
Fermi level versus pressure. The squares are the density of states
as a function of pressure, and the red line indicates the tendency
of density of state under the pressure. (\textbf{e}) T$_{c}$ is
plotted against pressure as well as evolution of MR, and
superconductivity appears at 12.8 GPa following the suppression of
XMR with increasing pressure. The BM transition appears after 16
GPa. The regions of XRM, SC, and BM in the phase diagram are
marked by blue, magenta, and green colors, respectively.  }
\end{figure*}

In previous studies, the emergence of SC is generally accompanied with a suppression of the XMR effect in topological SMs \cite{WTe2_DFKang_NatC,WTe2_XCPan_NatC,Pressure_PRL_LiSY}. There are different interpretations in the XMR suppression mechanism. Cai \textit{et al.} suggest that pressure weaken the electron-hole compensation in WTe$_{2}$ by gradually decreasing the hole carrier population \cite{Pressure_PRL_LiSY}. In-situ pressure-dependent Hall measurements by Kang \textit{et al.} indeed observe the change from a positive sign to negative sign in the Hall coefficients \cite{WTe2_DFKang_NatC}, suggesting increasing population of the electron carriers and decreasing hole carriers density. Moreover, Pan \textit{et al.} report that the SC transition is induced by rapid increase in the density of states at the Fermi surface, as a result of the compression of the unit cell \cite{WTe2_XCPan_NatC}, while the difference between hole pockets and electron pockets is enhanced with applied pressure.

NbAs$_{2}$ is a well compensated semimetal at ambient pressure
\cite{NbAs2family_YupengLi_arXiv}. When high pressure is applied,
XMR at 8 T is effectively suppressed from 5.16 at 0.5 GPa to 0.08
at 14.6 GPa, as shown in Fig.\ref{Figure2}c. The dwindling of XMR
as a function of pressure can be explained well by
pressure-induced decrease in the average mobility $\bar{\mu}$
\cite{FS_PRL_ZhuZW}, using the power-law relation of
$\text{MR}=(\bar{\mu}B)^{l}$. Using logarithmic scale, it is clear
that the XMR curves at different pressures have nearly the same
linear slope ($l=1.43$), which is a simple measure of the
electron-hole compensation accuracy \cite{TaSb2_WZ_Lifshitz}, and
the electron-hole compensation here is not so perfect. By fitting the
experimental data to the aforementioned equation, we found that
$\bar{\mu}$ decreases quasi-linearly as a function of pressure,
and only shows a slight upward curvature as entering the SC region
above 10 GPa (see Fig.\ref{Figure2}d). Our results thus imply that
a distinct phase boundary does not exist between SC and XMR, and
the mechanism of SC in NbAs$_{2}$ may be different compared to
other topological semimetals \cite{WTe2_DFKang_NatC}.

\begin{figure*}[!thb]
\begin{center}
\includegraphics[width=7in]{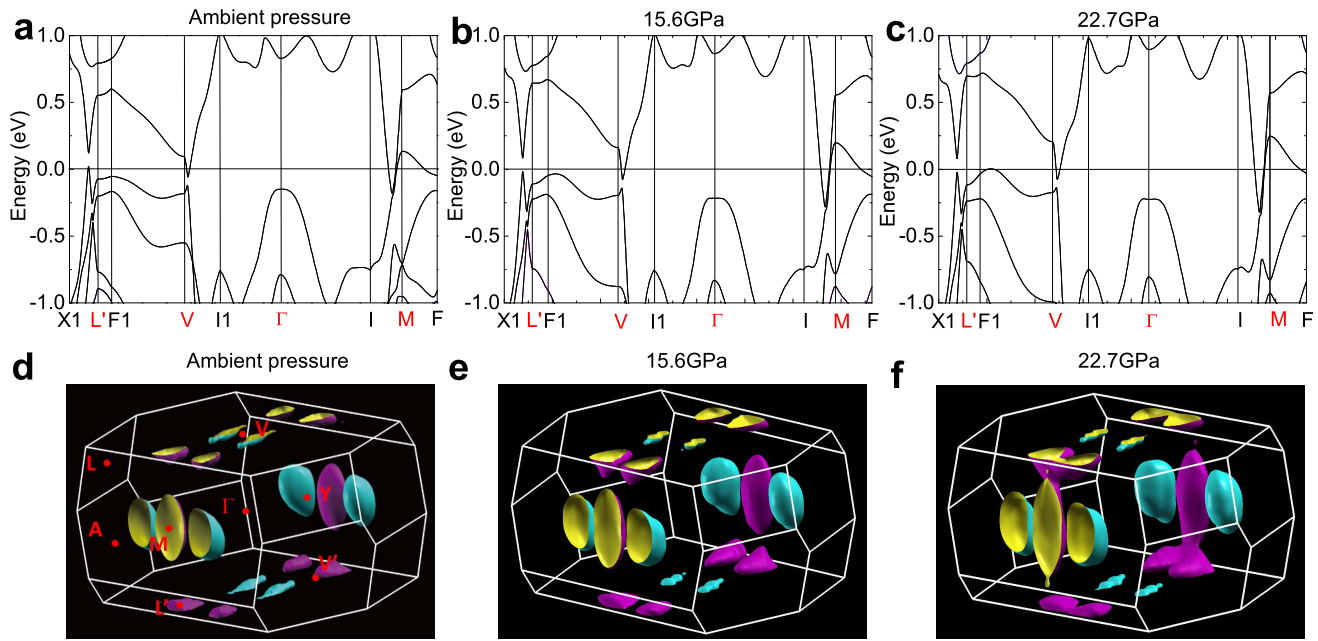}
\end{center}
\caption{\label{Fig4} Pressure-dependent band structures of NbAs$_{2}$, showing increasing electron and hole pockets. (\textbf{a, b and c}) Band structures in NbAs$_{2}$ are shown at ambient pressure, 15.6 GPa and 22.7 GPa, respectively. The red points are TRIM points. (\textbf{d, e and f}) 3D Fermi surface diagram at ambient pressure, 15.6 GPa and 22.7 GPa indicates that the TRIM (M) is enclosed by the hole-type Fermi surface (the magenta pockets), and blue pockets are electron-type Fermi surface. TRIM points are marked in the Fig.4d.  }
\end{figure*}

The positive resistance kinks, which become broader and shift to lower temperature as a function of pressure (\textit{P}), may be rooted in different mechanisms, such as superconductor to metal transition (SMT) in disordered system \cite{Griffiths_WJ_Science16}, a probable foreshadow of p-wave superconductivity in superconductor-ferromagnet nanowires structure \cite{SCandFM_WangJ_NP2010}, or highly anisotropic superconducting gap \cite{SrBi2Se3_WenHH_TSC}, but should not be the same as the semiconductor-like behaviour in $\text{LaO}_{0.5}\text{F}_{0.5}\text{BiSe}_{2}$\cite{LaO0.5F0.5BiSe2HP_LiuJZ_PRB}.  When external magnetic field (\textit{B}) is applied, $T_\text{c}$ monotonically shifts to lower temperatures and the positive resistance kinks gradually become predominant. As shown in Fig.\ref{Figure 3}a, it is surprising that the superconducting transition at 16 GPa is completely suppressed at 1000 Oe. This may be due to the predominant resistivity upturn, which pushes $T_\text{c}$ to below 1.8 K. Nevertheless, with the available experimental data, we can define an empirical critical $T_\text{c}^{90\%}$ by those points at which $R$ is $90\%$ of the normal state, as seen in inset of Fig.\ref{Figure 3}b. In Fig.\ref{Figure 3}b, the deduced $T_\text{c}^{90\%}$ can be well fitted by the Werthamer-Helfand-Hohenberg (WHH) model\cite{WHH_WNR_PR} , which yields the orbital-limited upper critical field in dirty-limited system by $H_{c2}^{orb}(0)\simeq-0.69\,T_{c}\times dH_{c2}/dT|_{T=T_\text{c}}=2288$ Oe.

Interestingly, by plotting the pressure-dependent $R$ at different
$T$ (4.5 K, 100 K, 200 K and 300 K, respectively) in
Fig.\ref{Figure 3}c, we are able to clearly see an anomaly at 12.8
GPa, which corresponds to the emergence of superconductivity.
However, there is no structural phase transition up to 29.8 GPa,
so the origin of this behavior needs further investigation. By
including the characteristic parameters of XMR,
T$_\text{c}^{90\%}$ as a function of pressure, we can draw the
$P-T$ phase diagram for NbAs$_{2}$. As summarised in
Fig.\ref{Figure 3}e, high pressure suppresses the XMR effect (the blue region) by
reducing the average mobility $\bar{\mu}$. The emergence of SC may
be related to the enhance of density of states
\cite{KTaO3_UenoK_NatNano2011,WTe2_XCPan_NatC}, and calculated
density of states increase as a function of pressure, as seen in
Fig.\ref{Figure 3}d. Moveover, there is a significant overlap
between the XMR (the blue region) and SC region (the magenta region) when superconductivity emerges at
12.8 GPa. Such a coexistence of XMR and SC is in contrast to the
other cases like WTe$_2$ where XMR is significantly suppressed as
entering the SC state under high pressure \cite{WTe2_DFKang_NatC}.
Above 16 GPa, there is a probable phase competition between SC and
the bad metal (BM) transition (the green region), which maybe lead to the vanishing
of $T_{c}$ at 27.9 GPa.

\vspace{3ex}

\noindent Electronic structures under pressure

\noindent To get insights into the aforementioned experimental results, we
have studied the pressure-dependent electronic structure of
NbAs$_{2}$ using density-functional theory (DFT) calculations,
which adopt the experimental lattice parameters determined by XRD.
Fig.\ref{Fig4} displays the DFT calculations at ambient pressure,
15.6 GPa, and 22.7 GPa, respectively. In excellent agreement with
the XMR measurements, the DFT results confirm that pressure
simultaneously increases electron and hole populations in
NbAs$_{2}$, which is displayed in Fig.\ref{Fig4}d,e,f where the
electron pockets (magenta) and hole pockets (blue) all enlarge
with increasing pressure. This is not surprising since the
monoclinic lattice is retained under high pressure. Equally
important, the steady growth in charge carrier does not show any
anomaly, neither in density of states in Fig.\ref{Figure 3}d nor
in Fermi surface topology, in the vicinity of 12.8 GPa when SC
starts to emerge. According to the proposed theory by Fu
\textit{et al.}, time-reversal-invariant topological
superconductivity in a centrosymmetry system requires odd-parity
spin pairing and an Fermi surface enclosing an odd number of TRIM
points \cite{CuxBi2Se3_FuL_PRL}. Although the pairing symmetry is
not explicitly known, the latter condition is well satisfied in
NbAs$_{2}$. As shown in Fig.\ref{Fig4}, the main hole pocket of
NbAs$_{2}$ enlarges under pressure, centering the TRIM point of
$M$. In addition, the electronic specific-heat coefficient of
$\gamma=0.45 mJ/mol/K^{2}$ is observed from the specific heat
measurements (see Fig.S3 in Supplementary Materials), suggesting
weak electronic correlation in NbAs$_{2}$ at ambient pressure.
With increasing pressure, the sudden hardening and broadening
phonon modes may start to play a critical role in the emergence of
SC, and the phonon-mediated pairing maybe have an odd-parity
symmetry in NbAs$_{2}$ to possess the singular behavior of the
electron-phonon interaction at long wavelength
\cite{turning_wanxg_NC,CuxBi2Se3PhonoMP_BPMR_PRB}. In addition, no
structural transition is observed as entering the superconducting
states with increasing pressure and thus the topological surface
state should remain undisturbed at high pressure, which suggests
the possibility of topologically superconducting surface states in
NbAs$_{2}$ because of superconducting proximity effect
\cite{Proximity_FuLiangPRL08}. Therefore, NbAs$_{2}$ with
time-reversal and inversion symmetries may be a candidate for
topological superconductor. It is necessary to be checked by
further experiments.

In summary, we study the high pressure-induced superconductivity in topological semimetal NbAs$_{2}$, which shows a superconducting transition temperature of 2.63 K at 12.8 GPa. Unlike previously reported topological semimetals, no structural phase transition occurs near the superconducting region, which is supported both by the high-pressure synchrotron X-ray diffraction and Raman data. Raman spectroscopy, resistance and specific heat data all suggest the critical importance of electron-phonon interactions on the SC pairing in NbAs$_{2}$, which may be a candidate of time-reversal-invariant topological superconductor with odd spin paring parity. Strikingly, the SC phase emerges with a nearly invariable electron-hole compensation and shows no anomaly in the density of states in the vicinity of the critical pressure of 12.8 GPa. Our results illustrate that high-pressure induced SC may be more complex in physical origins than the prevailing explanations, and the monoclinic family of MPn$_{2}$ may provide a platform to test various theoretical proposals and to search for topological superconductivity.

\vspace{3ex}

\noindent\textbf{METHODS}

\noindent Samples and experimental technique

\noindent Single crystals of NbAs$_{2}$ were grown by means of a vapour transport technique\cite{NbAs2family_YupengLi_arXiv}. High pressure was generated by a screw-pressure-type DAC consisting of nonmagnetic Cu-Be alloy and two diamonds with the culet of 300 $\mu$m diameter. A T301 stainless-steel gasket with a 280 $\mu$m diameter hole was used for different runs of high-pressure resistance measurement by the standard four-probe method. The single crystal was placed in the hole and a mixture of fine cubic boron nitride (CBN) powder with epoxy was compressed firmly to insulate the electrodes from the gasket. Pressure medium was daphne 7373 oil and some ruby powder was applied to demarcate the pressure by the ruby fluorescence method at room temperature. High-pressure synchrotron powder XRD ($\lambda = 0.4246\AA$) was performed at room temperature at the beamline 16 BM-D\cite{park2015new}, High Pressure Collaborative Access Team (HPCAT).

\vspace{3ex}

\noindent DFT calculations

\noindent The electron structure calculations were performed with the Vienna \textit{ab initio} simulation package (VASP) \cite{VASP_Kresse_PRB93,VASP_Kresse_PRB96} by the method of the projector augmented wave\cite{DFT_Blochl_PRB94} and the generalized gradient approximation (GGA)\cite{GGA_Perdew_PRL96} in order to introduce the exchange-correlation potential. Spin-orbit coupling has been included using the second variation perturbation method\cite{soc}. In addition, the plane-wave cutoff energy is setting about 500 eV and 21*21*13 k-points sampling is performed based on the Monkhorst-Pack scheme\cite{MPscheme_Monkhorst_PRB76}. The total energy is ensured to be converged within $10^{-6}$ eV. We use the structure parameters of high-pressure synchrotron X-ray diffraction to relax the structure with the tolerance of 0.01eV/\AA.s

~\\
\noindent Data availability

\noindent The data that support the findings of this study are available from the corresponding author upon reasonable request.

\vspace{3ex}

\noindent\textbf{ACKNOWLEDGEMENTS}

\noindent This work was supported by the National Key R\&D Program of China (Grant Nos. 2016YFA0300204, 2016YFA0300402, and 2017YFA0303002), the National Basic Research Program of China (Grant No. 2014CB921203), and the National Science Foundation of China (Grant Nos. U1332209, 11574264, 11774305). Y.Z. acknowledges the start funding support from the Thousand Talents Plan. The X-ray work was performed at HPCAT (Sector 16), Advanced Photon Source, Argonne National Laboratory. HPCAT operations are supported by DOE-NNSA under Award No. DE-NA0001974 and DOE-BES under Award No. DE-FG02-99ER45775, with partial instrumentation funding by NSF. The Advanced Photon Source is a US Department of Energy (DOE) Office of Science User Facility operated for the DOE Office of Science by Argonne National Laboratory under Contract No. DE-AC02-06CH11357.

\vspace{3ex}

\noindent\textbf{ADDITIONAL INFORMATION}

\noindent\textbf{Supplementary information} accompanies the paper on the \emph{npj Quantum Materials} website.

\vspace{3ex}

\noindent\textbf{Competing interests:} The authors declare no competing interests.

\vspace{3ex}

\noindent\textbf{Publisher's note:} Springer Nature remains neutral with regard to jurisdictional claims
in published maps and institutional affiliations.

\vspace{3ex}

\noindent\textbf{AUTHOR CONTRIBUTIONS}

\noindent Y.P.Li synthesized and characterized the single crystals. C.An, Y.H.Zhou, and Z.R.Yang conducted the high-pressure transport measurements.  Y.Zhou, C.An and R.R.Zhang carried out the Raman experiments,  X.L.Chen and C.Y.Park performed the high-pressure synchrotron X-ray diffraction experiments. C.Q.Hua and Y.H.Lu did the DFT calculations. Y.P.Li, C.Q.Hua, Y.H.Lu, Z.R.Yang, Y.Zheng and Z.A.Xu wrote the paper. All the authors analysed the data and discussed the results. Y.P.Li, C. An and C.Q.Hua were co-first authors to this work. Z.R.Yang and Z.A.Xu co-supervised the project.


\end{document}